\documentclass[12pt,a4paper]{article}
\usepackage[utf8]{inputenc}	
\usepackage{amsmath,amsthm,amsfonts,amssymb,amscd}
\usepackage{multirow,booktabs}
\usepackage[table]{xcolor}
\usepackage{fullpage}
\usepackage{lastpage}
\usepackage{enumitem}
\usepackage{fancyhdr}
\usepackage{mathrsfs}
\usepackage{wrapfig}
\usepackage{setspace}
\usepackage{calc}
\usepackage{multicol}
\usepackage{cancel}
\usepackage[retainorgcmds]{IEEEtrantools}
\usepackage[margin=3cm]{geometry}
\usepackage{amsmath}

\usepackage{empheq}
\usepackage{framed}
\usepackage[most]{tcolorbox}
\usepackage{xcolor}
\colorlet{shadecolor}{orange!15}
\usepackage{titling}

\begin{document}

\def\ra{{\rightarrow}}
\def\a{{\alpha}}
\def\b{{\beta}}
\def\l{{\lambda}}
\def\eps{{\epsilon}}
\def\T{{\Theta}}
\def\t{{\theta}}
\def\co{{\cal O}}
\def\car{{\cal R}}
\def\caf{{\cal F}}
\def\cs{{\Theta_S}}
\def\pr{{\partial}}
\def\tri{{\triangle}}
\def\na{{\nabla }}
\def\S{{\Sigma}}
\def\s{{\sigma}}
\def\sp{\vspace{.04in}}
\def\hs{\hspace{.25in}}

\newcommand{\be}{\begin{equation}} \newcommand{\ee}{\end{equation}}
\newcommand{\bea}{\begin{eqnarray}}\newcommand{\eea}
{\end{eqnarray}}

\begin{titlepage}
\vspace*{\fill}
\begin{center}
      {\Huge \scshape Perspectives of Perihelion Precession in Torsion Modified Gravity}\\[0.5cm]
      {\large R. Nitish, Rohit K. Gupta \& Supriya Kar}\\[0.4cm]
      {\large \slshape Department of Physics and Astrophysics,\\ University of Delhi 110007, India}\\[0.4cm]
    \end{center}
 \begin{abstract}
        Killing symmetries are revisited in $d$$=$$5$ bulk geometric torsion (GT) perturbation theory to investigate the perihelion precession. Computation reveals a non-perturbative (NP) modification to the precession known in General Relativity (GR). Remarkably the analysis re-assures our proposed holographic correspondence between a perturbative GT in bulk and a boundary GR coupled to $B_{2}$$ \wedge $$F_{2}$. In fact  the topological correction is sourced by a non-Newtonian potential in GR and we identify it with an ``electro-gravito'' (EG)  dipole. Interestingly the dipole correction is shown to possess its origin in a $4$-form underlying a propagating GT and leads to a NP gravity in $d$$=$$4$.
    \end{abstract}
    \vspace*{\fill}
\end{titlepage}
   

\section{Introduction}
General relativity is an elegant classical field theoretic formulation in $(3+1)$-dimensions by Einstein. It is a  relativistic and an interacting metric $g_{\mu\nu}({\mathbf{x}},t)$ field theory which ensures ten independent components and two local degrees of freedom. Thus four constraints are sourced by the inherent symmetries in GR. The additional four constraints are given by the Einstein field equations corresponding to the $0\mu$ components and they may provided a theoretical tool to explore new physics. In particular the Killing symmetries ensure the Killing equations, $i.e.$ the Lie derivative ${\cal L}_{K}g_{\mu\nu}$$=$$0$, which in turn yields $\nabla_{(\mu}K_{\nu)}$$=$$0$. A Killing vector $K_{\mu}$ is associated with a conserved charge underlying the symmetry. For instance see ref \cite{misner2017gravitation}.

\sp
In principle the conserved quantity sources a (gravitational) potential which in turn may include a quantum correction to the GR. Thus the Killing symmetries play a significant tool to explore a quantum correction to the exact solution(s) in GR. However a potential in GR is not unique and hence is very different in nature from that in a gauge theory or in an electro-magnetic theory. For instance, all known exact solutions in GR are defined with different gravitational potentials. Thus one of the primary motives  in GR is to identify an appropriate gravitational potential which incorporates  a quantum phenomenon consistently. Analysis in this paper reveals that a non-Newtonian potential $V_{q}$  can be a  candidate to describe a non-perturbative quantum correction.   In particular $V_{q}$ is  shown to describe an ``electro-gravito" (EG)  dipole correction underlying a topological BF-term to the GR.  The EG dipole  incorporates a  small nonzero length scale. It is believed to  describe the  Heisenberg's uncertainty principle consistently.

\sp
In the context various gauge theoretic tools have been explored in the last two decades to address the quantum gravity phenomena \cite{wilczek1998, sugamoto2002-PTP,bern2010-PRL, anastasiou2018-PRL}. Interestingly a dynamical generation of fourth or extra space dimension has been argued in GR \cite{arkani2001-PRL}.  For a detailed development leading to search for quantum gravity see ref \cite{hehl1995}. Interestingly modified gravity, underlying the elegant symmetries in GR,  has been explored in the recent past \cite{iorio2015editorial,debono2016general,vishwakarma2016einstein}.

\sp
On the other hand a constraint BF-theory has been argued to incorporate local degrees of GR \cite{plebanski1977separation}. There,  a two-form theory is believed to describe the Einstein gravity.  BF term is believed to describe new classical and quantum insights in GR \cite{mielke2010-PLB,Celada2016}. In particular the BF term  was derived from a  five dimensional bulk gauge theory by two of the authors \cite{NDS-Physica} in a GT formalism \cite{JHEP-Abhishek}. Generically the local degrees of a massless two-form gauge theory  is equal to the  local degrees of the metric field in a lower dimension  in presence of a scalar field dynamics. The equivalence between the local degrees of a gauge theory and gravity theory has led  to propose a new bulk gauge theory and boundary gravity correspondence \cite{NDS-Physica}. This perspective may allow one to view GR,  along with a topological correction, as a boundary phenomenon of a bulk two-form gauge theory. Contrary to the established bulk AdS and boundary CFT duality  \cite{witten,maldacena}, the proposed two-form equivalence  describes the GR generically. This is due to a fact that the emergent gravity is closed and hence  correctly represents the inherent  interacting nature of GR.   

\section{Gravitational potential}
In this article we address a pertinent issue possibly sourced by the Killing symmetries to a gravitational potential. Our analysis reveals a hint for quantum correction, realized via a topological coupling, to the GR. Presumably it is believed to describe the perihelion advances in GR. Interestingly we identify the topological term in the potential, with an EG dipole correction. Nonetheless the dipole is shown to be sourced by a conserved charge in a bulk geometric torsion (GT) theory on $R^{1,1}$$ \otimes $$S_3$.  Formally GT in ref  \cite{JHEP-Abhishek} is defined with a $U(1)$  gauge invariant three form ${\cal H}_{3}$$=$$d^{\cal D}{\cal B}_{2}$. It may also be re-expressed as a gauge potential in perturbation theory defined with a covariant derivative  $\nabla_{\mu}$ which is ensured by $d^{\nabla}{\cal H}_{3}$$\neq $$0$.               

\sp 
We compute the precession of the perihelion  with a renewed perspective in GT perturbation theory defined with an emergent metric \cite{JHEP-Abhishek} and estimate the extra space dimension. Results provide an evidence for a correspondence between the bulk GT dynamics and a boundary GR phenomenon in presence of an  EG dipole. In fact an afresh idea leading to a bulk (two form) gauge theory and boundary (Einstein) gravity has already been discussed by two of the authors in a collaboration \cite{NDS-Physica}. Interestingly a propagating torsion underlying a modified gravity has recently been addressed \cite{nikiforova2018infrared}. Along the line of thought the perihelion precession has  been revisited  for a plausible correction \cite{sultana2012,kupryaev2018concerning}. In particular ref \cite{sultana2012} elegantly discusses the perihelion precession in conformal-Weyl gravity. It has been shown that the perihelion  advance angle recieves a negative contribution which is compared with the data form the perihelion shift observations. Furthermore ref \cite{kupryaev2018concerning} has shown a correction in the perihelion precession angle obtained in GR. Analysis has revealed that perihelion of mercury around sun computed in GR shifts along the direction of motion of mercury.  In contrary our non perturbative analysis in this paper retains the formal expression of perihelion precession in GR and hence provides an estimate for the extra fifth dimension in the framework.  

\sp
In particular we begin with a static and $S_2$-symmetric vacuum solution in GR. The line element is known to describe the Schwarzschild black hole: $ds^2$$=$$-f\ dt^{2}+f^{-1}dr^{2}+r^{2}(d\theta^2+\sin^2\theta d\phi^2)$, where $f(r)$$=$$(1-2G_NMr^{-1})$ and $r$$>$$2G_NM$. The Schwarzschild geometry is characterized by one time-like Killing vector $K^{\mu}$$=$$(1,0,0,0)$ and three Killing vectors underlying the $S_2$-symmetry. The later  describes the angular momentum ${\mathbf L}$ and its magnitude is a conserved Noether charge $Q$. It takes a form $\xi^{\mu}$$=$$(0,0,0,1)$ and reflects the translational symmetry in $\phi$-coordinate. 

\sp
In the context we revisit an elegant tool described by the isometries of a Schwarzschild metric. The constant of motion for a time-like geodesics ensures:
\vspace{-0.1in}
\be
g_{\mu\nu}{dx^{\mu}\over{d\lambda}}{dx^{\nu}\over{d\lambda}}=-1\ .\label{motion-1}
\ee 
The equation for a Schwarzschild geodesics is worked out in a planar limit. It describes one dimensional motion of a particle of unit mass  \cite{carroll2019spacetime}
\begin{flalign} 
&& &\Big(\frac{dr}{d\lambda}\Big)^{2}+V_{\rm eff} =  E^{2}\ ,& \\
{\rm where}&& V_{\rm eff}&=\left (1-\frac{2G_NM}{r}+\frac{Q^{2}}{r^{2}}-\frac{2G_NMQ^2}{r^3}\right )\ . &\label{motion-2}
\end{flalign}
The second term in $V_{\rm eff}$ corresponds to the Newtonian gravitational potential and is known to be sourced by a scalar field. The third term also ensures the Newtonian gravity due to the equation of motion of the vector field sourced by an electric (and/or magnetic) charge $Q$. However the fourth term does not correspond to the Newtonian gravity as the apparent conserved force does not satisfy the inverse-square gravitational law. Thus a  non-Newtonian potential $V_q$$=$$(2G_NMQ^2)r^{-3}$ in $V_{\rm eff}$ is sourced by the conservation of both energy $E$ and charge $Q$ but turns out to be insignificant for large $r$ geometries. However for small $r$, the $V_q$ can incorporate a vital (relativistic) correction to the geometry underlying a classical vacuum in GR. In fact $V_q$ is known to describe the observed perihelion precession of planet(s) in an approximately closed path around the Sun and generically for the precessing elliptical orbit around a star. It ensures that these  orbits are not perfect ellipses in GR. They may allow a further possibility to explore the study of gravitational orbit from an alternate formulation underlying the perspectives in GR. In fact the precession of perihelion is one of the three experimental tests of GR suggested by Einstein. It may suggest that the observed precession of perihelion  advance possibly validates an alternate gauge theoretic formulation such as bulk GT. 
In the article we investigate the perspective of perihelion in a modified gravity underlying a bulk GT. The modification to Einstein gravity was constructed by one of the author in a collaboration \cite{JHEP-Abhishek}. It was shown that the torsion connection consistently modifies the covariant derivative $\nabla_{\mu}$ in Einstein gravity to ${\cal D}_{\mu}$ which satisfies ${\cal D}_{\mu}g_{\nu\lambda}$$=$$0$.
\section{Topological coupling to GR}
Interestingly a natural exclusion of Newtonian gravity may consistently ensure a quantum gravity phenomenon. In principle a non-Newtonian potential can describe interacting non-point masses and hence incorporates a lower cut-off on the radial distance $r$. Thus Heisenberg's uncertainty principle may be invoked in a non-Newtonian gravity and hence a quantum correction can be enforced by $V_q$ into the exact $S_2$-symmetric geometries in GR. It further ensures the background independence of the potential $V_q$ on an equatorial plane and hence may be interpreted as a NP correction to GR. For a recent review on  NP gravity we refer to ref\cite{ambjorn2012-PR}.
Analysis shows that the Killing symmetries in GR may provide a remarkable clue to unfold a topological correction at least to the maximally symmetric family of black holes defined with more than one tensor field source. For instance, the second and third terms in $V_{\rm eff}$ ensure the Reissner-Nordstr$\ddot{\rm o}$m (RN) black hole. The idea may lead to believe for a topological correction to the exact solution(s) in GR and together they may be described by a NP gravity:
\bea
S={1\over{4}}\int d^4x{\sqrt{-g}}\left ({{\cal R}\over{4\pi G_N}} - F_{\mu\nu}^2 - {1\over{3l^2}}H_{\mu\nu\lambda}^2\right ) -\ {1\over{4\pi G_N}}\int B_2\wedge F_2\ ,\label{q-action}
\eea
where $F_{\mu\nu}$$=$$(\nabla_{\mu}A_{\nu}$$-$$\nabla_{\nu}A_{\mu})$, $H_{\mu\nu\lambda}$$=$$(\nabla_{\mu}B_{\nu\lambda} + {\rm cyclic})$ and
the coupling $l$$=$$[{\rm length}]$. The $BF$-term incorporates a topological coupling to the metric dynamics via one and two form gauge potentials. For a detailed study on $BF$-gravity see refs\cite{krasnov2009,mielke2010-PLB, Celada2016}. 
A consistent truncation of the generic action (\ref{q-action}) yields a topologically coupled Einstein-Maxwell theory. Interestingly the dynamical terms in the action may alternately be derived from the $d$$=$$5$ Einstein gravity on $S^1$. It re-ensures that a higher dimensional gravity can be a potential tool to address a quantum gravity phenomenon in a lower dimension. However the $BF$-coupling in the action (\ref{q-action}) underlying the quantum correction needs to be placed by hand as it cannot be derived from $d$$=$$5$ (Kaluza-Klein) gravity. 
\section{Bulk GT$\longleftrightarrow$Boundary GR}
The mentioned difficulty may be resolved with a proposed correspondence \cite{NDS-Physica} between a two form perturbation theory, underlying a conformal symmetry, in $d$$=$$6$ bulk and a boundary ${\rm AdS}_5$. The idea has been modelled in a $U(1)$ gauge theory described by a two form in presence of a background gravity. On $S_1$ there are two massless two forms and the gauge group becomes $U(1)$$ \otimes $$U(1)$ in the $d$$=$$5$ bulk. Now the two forms in the bulk on $R^{(1,1)}$$ \otimes $$S_3$ under the bulk GT/boundary GR correspondence 
is described with all terms in $V_{\rm eff}$ in eq(\ref{motion-2}). 

\sp
A generic correspondence between the GT in bulk and the boundary (Einstein) gravity phenomenon is based on a number of evidences. For instance the non-Riemannian space-time curvature tensor $K_{\mu\nu\lambda\rho}$, in absence of a propagating torsion in bulk, has been shown to share all the properties of the Riemannian $R_{\mu\nu\lambda\rho}$ under the interchange of its indices \cite{JHEP-Abhishek}. Recall that the space-time curvature is an observable and the potentials (metric and two form) are not. Thus an observer would not  distinguish between the Einstein gravity and its alternate formulation with two form(s) gauge theory. In fact the scalar curvature $K$ computed from a two form ansatz in the bulk GT theory precisely identifies with the expression for the scalar $(C_{\mu\nu\lambda\rho}C^{\mu\nu\lambda\rho})$ in GR for a static vacuum, where $C_{\mu\nu\lambda\rho}$= conformal-Weyl tensor. Generically the local degrees of a mass-less two form on $R^1\otimes S_{d}$ is precisely equal to that of a metric on $S_{d}$ where the radius of $S_d$ may be identified with a pseudo-scalar field $\chi$. A vacuum expectation  $<$$ \chi $$>$$=$$\chi_0$ decouples the boundary phenomenon from the bulk ($\Sigma$) theory as the GT forms a condensate,$\, i.e.\, {\cal F}_4$$=$$0$. 
Then the bulk/boundary correspondence ensures an equivalence between the bulk GT action and the boundary GR which couples to 
the $BF$-term: 
\bea
S_{\rm bulk}&=&{{-1}\over{12\lambda^2}}\int_{\Sigma}d^5y{\sqrt{-G}}\ \Big ({\cal H}_{\mu\nu\lambda}^2 + H_{\mu\nu\lambda}^2\Big )\nonumber \\
&\equiv& \int_{\partial\Sigma} {1\over{16\pi G_N}}\Big ( d^4x\ {{\sqrt{-g}}R} -\ 4\pi B_2\wedge F_2\Big )\ ,
\label{torsion-bulk}
\eea
where ${\lambda}^2\equiv l^3$. The gauge theoretic forms $H_3$$=$$dB_2$ and $F_2$$=$$dA_1$ are defined with $\nabla_{\mu}$ derivative. All of them are $U(1)$ gauge invariants under a variation of their respective form fields. The $B_{\mu\nu}$ and 
${\cal B}_{\mu\nu}$ remain covariantly constant respectively with the original covariant derivative $\nabla_{\mu}$ and the modified derivative 
${\cal D}_{\mu}$. Thus in case of a pure ${\cal D}_{\mu}$ derivative theory, $B_{\mu\nu}$ behaves as a background field (meaning non-dynamical) and similarly in an original theory, ${\cal B}_{\mu\nu}$ turns out to be a background. However both the two forms are dynamical in the bulk action  (\ref{torsion-bulk}). In the case the ${\cal H}_{\mu\nu\lambda}$ may be re-expressed perturbatively:
\be
{\cal D}_{\mu}{\cal B}_{\nu\lambda}={1\over2}\left ({H_{\mu\nu}{}}^{\alpha}{\cal B}_{\alpha\lambda} + {H_{\mu\lambda}{}}^{\alpha}{\cal B}_{\nu\alpha}\right )\ .\label{torsion-1}
\ee
The original theory breaks the $U(1)$ symmetry spontaneously and hence leads to a massive ${\cal B}_{\mu\nu}$ description. Thus a mass-less 
${\cal B}_{\mu\nu}$ in bulk action (\ref{torsion-bulk}) absorbs all the local degrees in $B_{\mu\nu}$ which may identify with three Goldstone bosons. Interestingly a massive ${\cal B}_2$ quantum has been argued to describe a graviton and an instanton \cite{PTEP-NS} in any dimension $d$$ \ge $$4$. Along the line a propagating torsion and its relevance to the graviton has been worked out in the past \cite{hojman1979propagating, nikiforova2018infrared}.

\sp
Furthermore the $U(1)$ gauge invariance of ${\cal H}_{\mu\nu\lambda}$ under the ${\cal B}_{\mu\nu}$ variation has been shown to be restored in presence of an emergent metric \cite{JHEP-Abhishek}. It is given by
\be
G_{\mu\nu}=(g_{\mu\nu}- l^2{{\cal H}_{\mu}{}}^{\alpha\beta}{\cal H}_{\alpha\beta\nu})\ . \label{emergent}
\ee
Thus GT is defined with a NP derivative ${\cal D}_{\mu}$ and may be viewed via an absorption of a gauge theoretic torsion connections ${H_{\mu\nu}{}}^{\lambda}$ in the background. It retains the $U(1)$ 
gauge invariance non-perturbatively meaning ${\cal D}_{\mu}$ is treated  without its underlying perturbation in $\nabla_{\mu}$. Nonetheless the 
spontaneously broken $U(1)$ gauge invariance is restored with an emergent notion of metric in GT. Generically the bulk GT dynamics is derived with a non-vanishing ${\cal F}_4$$=$$d{\cal H}_3$ which in turn governs an axionic field $\chi$. Then $[{\cal F}_4, {\cal H}_3, H_3]$ in bulk GT would describe a massive GT coupled to $B_{\mu\nu}$ dynamics. Under the bulk/boundary proposal the boundary GR would like to  correspond to $[{\cal R}$$-$$2\Lambda]$ for a cosmological constant $\Lambda $$< $$0$ and with a $BF$-correction. This in turn is believed to describe an 
${\rm AdS}_4^Q$ on a boundary theory with a NP correction \cite{NDS-Physica}.

\sp
Though the boundary theory is closed from the perturbative bulk perspective, it describes multiple ${\rm AdS}_4$ patches with local boundaries (within the closed hyper-surface) in NP gravity. Here the superscript $Q$ signifies a quantum or quintessence underlying an axion and sourced by the topological correction. We believe that the $BF$-term \cite {Celada2016}  possesses a deep implication to the Big Bang cosmology and its investigation may unfold the mysteries behind the dark energy in universe.  For a recent article on dark matter see \cite{Rosenberg2019}.
\section{Perihelion precession in bulk GT} 
A circular orbit in Newtonian gravity turns out to be Keplerian or elliptical due to the planar effect meaning the motion of other massive bodies in the same plane. However the relativistic effect in GR ensures that the perihelion advances along its azimuthal angle $\phi$$ \rightarrow $$\phi$$+$$2(\pi $$+$$\delta \phi)$. The precessed angle has been computed in GR to yield: $\delta \phi $$=$$ 6\pi(G^2_N M^2Q^{-2}) $$ > $$ 0$. It shows a deviation from a perfect elliptical orbit to an open path which signifies the role of GT in $d$$=$$5$ and for $d$$ \ge $$5$.  In a different context interesting precession computations  have been obtained in literature \cite{iorio2004possibility,adkins2007orbital,iorio2012constraining}.  Very recently a perturbative precession of the  mean anomaly at the epoch was unfolded   \cite{Iorio:2019rfz}. A detailed description on mean anomaly though interesting but  is beyond the scope of this article. 

\sp
Now we begin with an $U(1)$ gauge theory described by a dynamical $B_2$ and a covariantly constant ${\cal B}_2$ in $(4+1)$. 
The gauge ansatz \cite{JHEP-Abhishek} for positive constants ($b,P$) is given by 
\be
{\cal B}_{t\psi}=b={\cal B}_{R \psi}\qquad {\rm and}\qquad B_{\psi\phi}=P\sin^{2}\psi \cos\theta\ .\label{ansatz-2}
\ee 
The non-trivial components of GT becomes:
\begin{flalign}
&&{\cal H}_{\theta\phi\psi}&=bPl^{-1}\sin^{2}\psi\sin\theta &\nonumber\\
{\rm and}&& {\cal H}_{\theta\phi t}&={\cal H}_{\theta\phi R}=\frac{-bPl}{R^{2}}\sin^{2}\psi \sin\theta\ . &\label{ansatz-3}
\end{flalign}
Using an emergent metric (\ref{emergent}) a line-element has appropriately been approximated to yield $d$$=$$5$ Schwarzschild black hole \cite{JHEP-Abhishek}:
\be\label{schwarzschild-5d}
ds^{2}=-\left[1-{{2m}\over{R^2}}\right] dt^2+ \left[1-{{2m}\over{R^2}}\right]^{-1}dR^{2}+ R^2 d\Omega^2_3\ ,
\ee
where $2m$$=$$(bPl^2)$ and the $S_3$ line element $d\Omega^2_3$$=$$(d\psi^2 + \sin^2\psi [d\theta^2 +\sin^2\theta\ d\phi^2])$ with $0$$<$$\psi$$ \le $$ \pi $, $0$$<$$\theta$$ \le $$ \pi $ and $0$$<$$\phi$$ \le 2$$\pi$.  In the case the emergent Schwarzschild geometry in $d$$=$$5$ is characterized by one time-like Killing vector $K^{\mu}$$ \rightarrow $$(1,0,0,0,0)$ and six Killing vectors underlying the $S_3$. The translation symmetry in $\phi$ is characterized by $\xi^{\mu}$$ \rightarrow $$(0,0,0,0,1)$. On equator two conserved quantities are
$E$$\rightarrow $$(1-\frac{2m}{R^2}) \frac{dt}{d\lambda}$ and $Q$$\rightarrow $$R^2\frac{d\phi}{d\lambda}$. 

\vspace{0.06in}
The constant of motion (\ref{motion-1}) in $d$$=$$5$ on an equatorial plane for the time-like geodesics is worked out to yield: 
\vspace{-0.1in}
\be
\left ({{dR}\over{d\lambda}}\right )^2 + \left (1-\frac{n}{R^2}-\frac{q^2}{R^{4}}\right )={\cal E}\ , \label{Energy}
\ee
where $n$$=$$(2m-Q^2)$, $q^2$$=$${2mQ^2}$ are constants and ${\cal E}$$=$$E^2$ is an analogue of total energy. The second term within the bracket in (\ref{Energy}) corresponds to an effective potential in $d$$=$$5$.  Furthermore a second term $(n/R^{2})$ in $V(R)$ is an empirical generalization of Newtonian potential to $d$$=$$5$. Similarly the third term $(q^{2}/R^{4})$ in $V(R)$ is a generalization in $d$$=$$5$ from $q^{2}/r^{2}$ term in $d$$=$$4$. In fact potential $V(R)$ is derived appropriately  from eqn (\ref{motion-1}) and eqn (\ref{schwarzschild-5d}). Interestingly the non-Newtonian  term present in GR is absent in $d$$=$$5$ Einstein gravity.  This is indeed a striking difference between GR and GT. Analysis leading to a derived $V_{\rm eff}$ in GT and in GR may suggest  a hint for  quantum correction  to GR. In fact the $V_{\rm eff}$ in GT perturbation theory exploits an emerging notion of metric and hence both GT and GR are defined with a derivative $\nabla_{\mu}$ respectively in $d$$=$$5$ and $d$$=$$4$. 
Apparently the eq(\ref{Energy}) describes the $d$$=$$1$ motion of a unit mass (classical) particle in an effective potential ${1\over2}V(R)$ and with a total energy $({\cal E}/2)$. This unusual energy analogue equation signifies that the actual motion involves the motion of a planet around the sun and hence $t(\lambda)$ and $\phi(\lambda)$ do join the $R(\lambda)$ equation (\ref{Energy}). Thus the actual scenario is drastically different from that of $d$$=$$1$ motion of a particle. The expression for $Q^2$ is used to re-express the eq(\ref{Energy}) under a change of variable. It takes a form:
\be
D_5=\Big (\frac{dR}{d\phi}\Big )^{2}+(1-{\cal E})\frac{R^4}{Q^2}+\Big (1-\frac{2m}{Q^2}\Big )R^{2}-2m =0\ .\label{Energy-1}
\ee
If $w$ is a fourth space coordinate then $R^2$$=$$(r^2$$+$$w^2)$. For $w^{2}$$ \ll $$r^{2}$, the $R^4$$ \approx $$r^4( 1$$+$$2w^2r^{-2})$. Then the  eq(\ref{Energy-1}) may be reduced to that in $d$$=$$4$. However using $X$$=$$Q^2(mr)^{-1}$ the eq(\ref{Energy-1}) is re-expressed as:
\begin{flalign}
&&{{d^2X}\over{d\phi^2}} &- \left ({{3G_N^2M^2}\over{Q^2}}\right )X^2 + X=1 &\nonumber\\
{\rm and}&& (aX^4 &+ bX^3 + cX^2 +d)=0\ , &\label{Energy-3}\\
{\rm where}&& a&={{15m^2}\over{2Q^2}},\ b+4={{4m}\over{w^2}}+\frac{4 m}{Q^{2}}, &\nonumber \\
{\rm and}&& 3&-2c={{3Q^2}\over{w^2}}, d={{Q^4}\over{2( mw)^2}}. &
\end{flalign}
A generic solution $X^{(GT)}$$=$$(X^{(GR)} + X_5)$ can be approximated with $w^2$$ \ll $$Q^2$ and identifying $\alpha=(3G_N^2M^2)Q^{-2}$ as the azimuthal precession in GR. Then: 
\be
w^2\approx m\left (1+\frac{15}{8 \sqrt{3}}\frac{m}{Q}\right )\quad {\rm and}\;\ X_5\approx -\frac{4Q^2\alpha}{3w^2}\ .\label{extra}
\ee
Explicitly:
\bea
X^{(GT)}&=&1+e \cos\phi +e\alpha \phi \sin\phi+\alpha \tilde{\alpha}\ ,\; \tilde{\alpha}=\left (1-\frac{4Q^2}{3w^2}\right ),\nonumber\\
&=&1+e'\cos\phi+ e'' \alpha \phi\sin\phi\ ,\label{soln-1}
\eea
where 
$e'$$=$$(e+\alpha\tilde{\alpha}\cos\phi)$, $e''\phi$$=$$(e\phi+ \tilde{\alpha}\sin\phi)$ and $e$= eccentricity of ellipse. For $w^2\rightarrow w_0^2={{4Q^2}\over{3}}$, the $\tilde{\alpha}=0$ and $X_5$ contribution vanishes. Nevertheless for $\tilde{\alpha}\neq 0$:  
\bea
X^{(GT)}&=&1 +e' \cos\phi +e' \tilde{\alpha} \phi \sin \phi\ ,\ \nonumber\\
&=&1+e'\cos [(1-\tilde{\alpha})\phi]\ .\label{soln-2}
\eea
Interestingly the GT solution in a limit ${\tilde{\alpha}}$$ \rightarrow $$\alpha$ identifies with that in GR. This is due to a fact that 
$\alpha^2$ is insignificantly small. An estimate for $w$ is worked out for the motion of the planet Mercury around the Sun in torsion gravity. It yields $w_0=1.04\times 10^7 m$, where we have used $e=2\times 10^{-1}$, semi-major axis 
$a=5.8\times 10^{10}$m and $(G_NM_{\rm Sun}c^{-2})= 1.5\times 10^3$m. In the case an extra dimension turns out to be $10^3$ times smaller than the remaining three space dimensions which along with a time coordinate describes the GR. Thus a perihelion advances by $w_0$ in an orthogonal direction to the remaining $3$-space coordinates. A small elevation in periodicity of the azimuthal angle $\phi$ is along a resultant direction to $\omega$ and $\mathbf{r}$. It is due to the non-planar effect underlying an intrinsic (non-commutative) nature of rotations which are only possible off a plane. Generically the bulk GT theory describes a spiral path for a planet and hence an open path! It is due to a propagating (axionic) scalar $\chi$ along the $w$-direction. Thus an assigned vacuum expectation $\chi_0$ can fine tune $w_0$ to a smaller value! A small $w_0$ ensures a nearly closed elliptical orbit in GR. A spiral path followed by a planet in GT may be approximated to describe an elliptical orbit on a slanted plane in GR.
\section{Electro-gravito dipole}
We recall the potential $V_q$ for a plausible physical interpretation in GR. 
It is obvious to note that the BF-term in the proposed action (\ref{q-action}) does not modify the known exact geometries in GR. Interestingly the $V_q$ may be identified with an EG dipole term where $M$$\neq $$0$. The coined name presumably ensures the formation of an electric dipole only in presence of Einstein gravity. It may also be viewed through the coupling of vector field $A_{\mu}$ to the metric field $g_{\mu\nu}$ as in Einstein-Maxwell action. Thus two opposite electric charges $\pm Q$ are separated by $M$ and the EG  dipole is defined with a coupling $G_N$ which replaces the coulomb constant $(4\pi\epsilon_0)^{-1}$ in a typical electric dipole. However an electric dipole correction is ruled out in GR  due to a fact that electric charges are not sourced by the metric field. It goes as $r^{-3}$   and hence is insignificantly small for the large length scale in GR.  Nevertheless an electric quadruple can be configured with two EG dipoles placed closed to each other with opposite and equal charges. The net charges at the poles cancel each other to form a closed gravitational loop in a quadruple and hence is known to contribute a correction to the GR. 

\sp
An EG dipole in Einstein-Maxwell theory is argued to possess its origin in a $BF$-term. A topological number ensures the number of windings which in turn would like to describe the multi Reissner-Nordstr$\ddot{\rm o}$m (RN) black hole. Thus semi-classical vacua can be described with a quantum tunneling of an instanton via the $BF$-term underlying an EG dipole. The perspective of the $BF$-boundary term has been shown to be sourced by the bulk $B_2$ dynamics \cite{NDS-Physica}. In fact the $B_2$ ansatz in the bulk gauge theory under $S_3$$\rightarrow $$S_2$, $i.e.$ for the second polar angle $\psi$$\rightarrow $${{\pi}\over2}$, has been worked out in ref\cite{PRD-Abhishek}. Our result matches with the expression for $V_q$ which sources the experimentally observed perihelion precession of planets in the solar system. Analysis may compel to revisit the perihelion precession with a renewed perspective in a bulk GT on $R^{1,1}$$ \otimes $$S_3$. 

\sp
In the context we recall the two form ansatz (\ref{ansatz-2}) to construct a geometric torsion (\ref{ansatz-3}) in a  perturbation theory. Thus, the gauge invariance is spontaneously broken in perturbation theory. As a result the 
${\cal H}_3$ may be treated as a gauge potential to define an $U(1)$ gauge invariant ${\cal F}_4$$=$$d{\cal H}_3$$ \neq $$0$. This in turn describes a propagating pseudo scalar presumably sourcing a gravitational instanton in the bulk geometric torsion theory. The components of field strength for a dynamical GT are worked out using the gauge ansatz (\ref{ansatz-3}). They are given by
\bea
{\cal F}_{t\psi \theta \phi}&=&-{\cal F}_{R \psi \theta \phi} = \frac{-2bP}{R^{2}}\sin 2\psi \sin \theta\nonumber\\
{\cal F}_{t R \theta \phi}&=&\frac{2bPl}{R^{3}}\sin^{2}\psi \sin \theta\ .\label{4-form}
\eea
On an equatorial plane the non-zero component ${\cal F}_{t R\theta \phi}$$=$$(2bPl)R^{-3}$ precisely identifies with that of an EG dipole term in an effective potential (\ref{motion-2}) obtained in GR as $R$$\rightarrow $$r$ due to a fact that the spherical symmetry becomes insignificant on a plane. However the $3$-form equations of motion $\nabla^{\mu}{\cal F}_{\mu\nu\lambda\rho}$$=$$0$ in the perturbation GT theory ensures that the $4$-form ansatz does not contribute to the Newtonian force. It leads to a consistent geometric description in $d$$=$$5$. Thus the $4$-form ansatz re-confirms a holographic correspondence between a classical bulk (perturbation GT) and a boundary GR (with a NP correction). It is consistent with the idea of ${\rm AdS}_5/{\rm CFT}_4$ correspondence \cite{witten,maldacena} which maps a weakly coupled bulk to a strongly coupled boundary. 

\sp
A propagating GT and the metric in GR respectively require a minimal $d$$=$$5$ and $d$$=$$4$. In fact an emergent metric $G_{\mu\nu}$ has been shown to restore gauge invariance in GT theory \cite{JHEP-Abhishek}. Nonetheless the NP term in the action uses a modified derivative ${\cal D}_{\mu}$ and the $U(1)$ gauge invariance is maintained in GT by definition. It has been argued that the graviton in $d$$=$$4$ may equivalently be described by the quantum of a massive two form \cite{PTEP-NS}. 
\section{Concluding remarks}
The observed perihelion precession, experimentally validating the GR, is believed to describe a nearly elliptical orbit for a planetary motion. In fact the advanced azimuthal angle, from the perspective of an equatorial plane in GR, has lead to an open path which in turn is approximated as an elliptical orbit on a slanted plane. In the context we revisited the phenomenon in bulk GT formulation where an instanton, in the disguise of a propagating torsion, was shown to incorporate the advances in perihelion along a spiral path. Remarkably analysis reveals an evidence to the proposed correspondence between a  bulk GT/boundary GR in presence of a $BF$-term \cite{NDS-Physica}. The topological correction was identified with an EG dipole in GR and it is  believed to describe a shade of quantum gravity possibly leading to a degenerate (multi) RN black hole vacua.

\sp
Furthermore it would be interesting to explore the perturbative precession  computation  in an alternate formulation of  gravity underlying teleparallelism \cite{muller1983teleparallelism,mielke2017teleparallelism}. A torsion is solely  believed to govern the space-time curvature in teleparallel theory and hence it may be viewed as a  (Einstein) gravity decoupling limit of quantum gravity. Our analysis has revealed that (geometric)  torsion gravity plays a significant role in a higher space dimension than GR. Intuitively an extra fifth dimension incorporates a non zero  length scale in a torison gravity which in turn is responsible for a quantum description of gravity consistently. Interestingly the length scale makes an open path for a planetary motion which is approximated as an elliptical orbit in GR. Thus we believe that torsion gravity may play a vital role to explain some of the quantum gravity phenomena.

\vspace{0.7in}


\end{document}